\documentclass[twocolumn,showpacs,preprintnumbers,amsmath,amssymb]{revtex4-1}
\usepackage{graphicx}
\usepackage{dcolumn}
\usepackage{bm}
\usepackage{color}
\begin{document}
%
%
\title{Strain gradient induced electric polarization in $\alpha$-phase polyvinylidene fluoride films under bending conditions}
\author{Sivapalan Baskaran and Xiangtong He\\\small\emph{Department of Mechanical and Aerospace Engineering,\\The State University of New York, Buffalo, New York 14260, USA}\\\ \\Yu Wang\\\small\emph{Department of Materials Science and Engineering, Michigan Technological University, Houghton, MI 49931, USA}\\\ \\John Y. Fu}
\email{youweifu@buffalo.edu}
\affiliation{Department of Mechanical and Aerospace Engineering,
The State University of New York, Buffalo, New York, 14260, USA}
%
%
%
%
%
\begin{abstract}
The relationship between the applied elastic strain gradient and the induced electric polarization in the $\alpha$-phase polyvinylidene fluoride (PVDF) films under bending conditions has been investigated. Our experimental studies have shown that the flexoelectric polarization is linearly proportional to the strain gradient and the corresponding direct flexoelectric response is strong. It is reasonable to believe that the physical mechanisms behind the flexoelectric effects in polymers and solid dielectrics are different.
\end{abstract}
\maketitle
%
%
The flexoelectric effect, which describes the linear energy coupling between the applied strain gradient and the induced electric polarization in solid crystalline materials, was proposed by Kogan in 1964 \cite{kogan1964}. Kogan's model can be written as
\begin{equation}
P_{i}=\mu_{ijkl}\nabla_{l}S_{jk},
\label{flexo1}
\end{equation}
here $P_{i}$ is the induced polarization; $\nabla_{l}$ is the symbol representing the gradient with respect to the direction $l$, $\mu_{ijkl}$ represents the flexoelectric coefficient, and $S_{jk}$ is the applied strain. Initially, Kogan used the term of ``{\it piezoelectricity}" to describe this new phenomenon in his paper \cite{kogan1964}. Indenbom {\it et al.} later suggested using the term of ``{\it flexoelectricity}" to define the effect since the physical mechanisms behind piezoelectricity and flexoelectricity are different \cite{indenbom1981}. The term, ``{\it flexoelectricity}", was coined by de Gennes for the description of an analogous effect in liquid crystals \cite{degennes1974}.

The physical picture of the flexoelectricity in solid crystalline dielectrics is clear; it is the inhomogeneous strain that breaks the inversion symmetry and induces the electric polarization even in centrosymmetric crystalline dielectrics \cite{sharma2007}. The flexoelectric coefficient, $\mu_{ijkl}$, was previously estimated to have an order of magnitude of $e/a$ ($e$ and $a$ represent the electron charge and the atomic dimension of the unit cell of crystalline dielectrics, respectively), which is approximately equal to $\mathrm{10^{-10} C/m}$ (see Ref.\cite{kogan1964} and references therein). Tagantsev further pointed out that this coefficient should be proportional to electric susceptibility, which can be written as \cite{tagantsev1985,tagantsev1986,tagantsev1991}
\begin{equation}
\mu_{ijkl}=\chi_{ij}\gamma_{kl}(\frac{e}{a}),
\label{flexo2}
\end{equation}
here $\chi_{ij}$ is the susceptibility of bulk dielectrics under the mean field approximation, $\gamma_{kl}$ is an inhomogeneous susceptibility distribution tensor.

The flexoelectric phenomena in liquid crystals have also been extensively studied over the past several decades. In 1969, Meyer proposed his phenomenological model to describe an analogous flexoelectric effect in liquid crystals, i.e., the linear energy coupling between the $\mathit{splay}$ and $\mathit{bend}$ directing deformations and the induced electric polarization, which can be written as \cite{meyer1969}
\begin{equation}
P_{d}=e_{11} \, (\vec{n} \, \mathbf{div} \, \vec{n})+e_{33} \, (\vec{n} \, \times \mathbf{curl} \, \vec{n}),
\label{flexo3}
\end{equation}
here $\vec{n}$ is defined as the director that represents the orientational deformation in liquid crystals, $P_{d}$ represents the induced electric polarization along the direction of $\vec{n}$, and $e_{11}$ and $e_{33}$ denote the flexoelectric coefficients related to $splay$ and $bend$ distortions in liquid crystals, respectively. Like Kogan, Meyer also used the term of ``{\it piezoelectricity}" in his model when he published his paper. Largely based on de Gennes's work \cite{degennes1974}, we now call such phenomena in both solid and liquid states as ``{\it flexoelectricity}" or ``{\it flexoelectric effect}".

The reason that the inhomogeneous deformation and the corresponding potential energy could exist in liquid crystals and are not destroyed by molecule diffusing in the liquid state is that liquid crystals are viscous. It is the viscosity that helps liquid crystals preserve the aforementioned inhomogeneous deformation, which renders the existence of flexoelectricity in liquid crystals. Thus the physical mechanism of the flexoelectric phenomena in liquid crystals is more complicated than that in solid crystalline dielectrics. It is well known that the flexoelectricity in liquid crystals is closely related to the geometrical asymmetry of the mesogen molecules or $\mathit{shape \, polarity}$, but the relationships between the flexoelectric coefficients, $e_{11}$ and $e_{33}$, and molecular structures and between the flexoelectricity and the viscosity are far from being understood.

\begin{figure}[h!]
\begin{center}
\includegraphics[width=1.0\columnwidth]{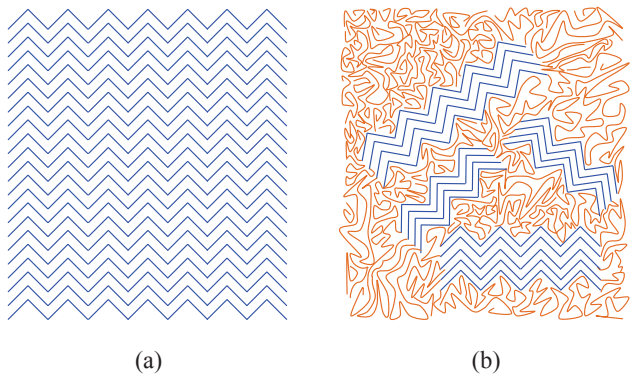}
\caption{(a) a perfect crystal (we use a zig-zag pattern to diagrammatically represent the crystalline structure); (b) a partially crystallized polymer with both the crystalline structure and the amorphous structure.}
\label{fig1}
\vspace{-0.25in}
\end{center}
\end{figure}

The studies of the flexoelectricity in polymers are also challenging since the flexoelectric responses of polymers must involve certain time-dependent molecular structure changes. The fundamental difference between polymers (including solid polymer films) and solid crystalline materials resides in the inherent rheological or viscoelastic properties of polymers, i.e., polymers possess both viscosity and elasticity simultaneously. Therefore, the long-term flexoelectric responses of polymers should also possess rheological behavior. Besides, the short-term flexoelectric responses of polymers are dependent on not only their molecule rotation but also the interaction between their different phases, which distinguishes the flexoelectric behavior in polymers from that in both solid crystalline dielectrics and liquid crystals. We use a diagrammatic drawing to further demonstrate this viewpoint. In Fig. 1, we can see the difference between a perfect crystal structure (shown in Fig. 1(a)) and a semi-crystalline structure (shown in Fig. 1 (b)); a polymer with the latter structure can be regarded as a ``{\it spontaneous composite}" formed with crystalline parts in a glassy matrix \cite{degennes1990}, which possesses some unique properties. When such a semi-crystalline polymer is undergoing deformation, all of its parts (both crystalline and amorphous ones) will make contribution to its flexoelectric responses. However, the net flexoelectricity of the polymer is not the sum of that of its different parts or phases; the interaction between those phases might dominate its flexoelectric responses under certain circumstances, which has been partially confirmed by our recent experimental observations \cite{fu2011a,fu2011b}. Therefore, the theoretical models and experimental techniques developed for the flexoelectricity studies of solid crystalline materials and liquid crystals might not be appropriate for that of polymers. Unfortunately, only a few experimental results about the flexoelectric properties of polymers have been reported so far; our understanding of the flexoelectric behavior in polymers is still rather limited.

\begin{figure}[h!]
\begin{center}
\includegraphics[width=1.0\columnwidth]{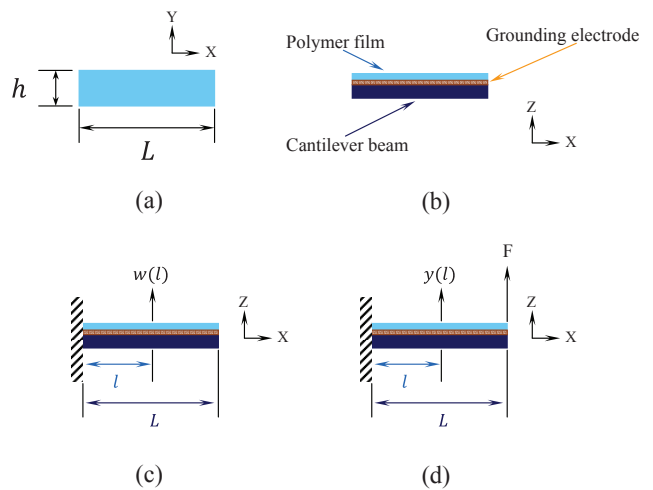}
\caption{(a) a rectangle shape PVDF film with $h$ = 20mm and $L$ = 55mm, the thickness of the film is 13.5$\mathrm{\mu}$m; (b) the cross-section
view of an effective beam system (the rectangle shape PVDF film + a cantilever beam); the grounding electrode is fabricated on the bottom surface of the PVDF film, and then the film is glued by using high strength epoxy to the top surface of the cantilever beam; (c) the deflection, $w(l)$, of the clamped effective beam system undergoing free vibration, $l$ = 11.5mm and $L$ = 55mm; (d) the static deflection, $y(l)$, of the clamped effective beam system under an end load F, $l$ = 11.5mm and $L$ = 55mm.}
\label{fig2}
\vspace{-0.25in}
\end{center}
\end{figure}

As shown in Eq.\ (\ref{flexo2}), the flexoelectric coefficient is linearly proportional to the electric susceptibility. Such a relationship has been proposed to exist in polymers too \cite{marvan1988,marvan1991,marvan1997}. Previous experimental studies also showed that the order of magnitude of the induced polarization to the deformation gradient ratio was approximately $\mathrm{10^{-11}C/m}$ in elastomers \cite{marvan1988}, which is in good agreement with the theoretical estimation mentioned, for example, in Kogan's paper \cite{kogan1964}. However, J\'{a}kli and his co-workers recently observed a giant flexoelectric phenomenon in certain bent-core liquid crystal elastomers (BCLCEs) with the nematic phase \cite{harden2010}, which contradicts the belief that the flexoelectric coupling is small in polymers. Prompted by their work, we investigated the flexoelectric properties of both quasi-amorphous and semi-crystalline polyvinylidene fluoride (PVDF) films under tensile stretching conditions; the giant flexoelectric phenomena and other interesting material behavior have also been observed in those films \cite{fu2011a,fu2011b}. These experimental observations might indicate that the Tagantsev's model given in Eq.\ (\ref{flexo2}) might not be applied to the flexoelectric behavior of polymers.

In this Letter, we investigate the flexoelectricity of semi-crystalline $\alpha$-phase PVDF films under bending conditions. PVDF is a prototypical semi-crystalline polymer, which has several forms, such as the apolar $\alpha$ phase, the polar $\beta$ phase, etc. The $\alpha$ phase of PVDF has the crystallographic point group symmetry 2/m ($\mathrm{C_{2h}}$ in the Schoenflies notation), whereas the $\beta$ phase of PVDF has the point group symmetry mm2 ($\mathrm{C_{2v}}$ in the Schoenflies notation). The method of preparation of the $\alpha$-phase PVDF film and its XRD characterization have been given in Ref.\ \cite{fu2011b} and will not be repeated here.

The $\alpha$-phase PVDF films are cut into the pieces with the rectangle shape shown in Fig. 2(a). The grounding electrode is fabricated on the bottom surface of a rectangle shape PVDF film, and then the film is glued by using high strength epoxy to the top surface of a cantilever beam to form an effective beam system shown in Fig. 2(b). The direct flexoelectric effect is measured from the effective beam system undergoing inhomogeneous deformation generated by its bending movement. Using a bending method to measure the flexoelectric behavior of solid dielectrics has been extensively studied by Ma and Cross (for simplicity, we shall use ``Ma-Cross method" to represent their bending method from now on) \cite{ma2001a,ma2001b,ma2002,ma2003,ma2005}. The Ma-Cross method can be summarized as follows: assume an effective beam system, shown in Fig. 2(c), is undergoing a natural or free vibration, its deflection $w(l)$ can be written as \cite{meirovitch2001}
\begin{widetext}
\begin{equation}
w(l)=A_{1}\left\{\left[sin(\beta L)-sinh(\beta L)\right]\left[sin(\beta l)-sinh(\beta l)\right]+\left[cos(\beta L)+cosh(\beta L)\right]\left[cos(\beta l)-cosh(\beta l)\right]\right\},
\label{flexo4}
\end{equation}
\end{widetext}
where $A_{1}=C/[sin(\beta L)-sinh(\beta L)]$, and we here only consider the fundamental mode of the vibration. Therefore, we have $\beta L=1.875$ \cite{meirovitch2001}. $C$ can be determined from the boundary condition, i.e., the measured displacement in the Z axis direction of the effective beam system. Both the direction of $w(l)$ and the Z axis are shown in Fig. 2(c).

\begin{figure}[h!]
\begin{center}
\includegraphics[width=1.0\columnwidth]{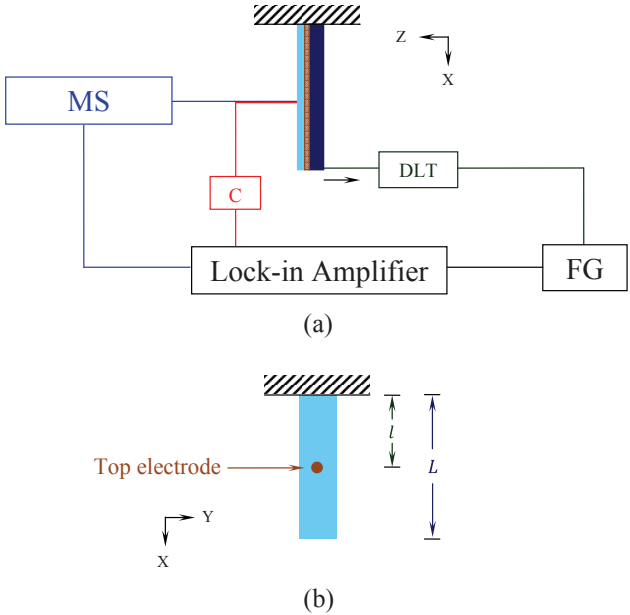}
\caption{(a) a lock-in detection setup; C: current measurement, FG: function generator, MS: MicroStrain DVRT probing sensor, DLT: dynamic loading tester; (b) the top electrode is fabricated on the top surface of the PVDF film; $l$ = 11.5mm and $L$ = 55mm.}
\label{fig3}
\vspace{-0.20in}
\end{center}
\end{figure}
\begin{figure}[h!]
\begin{center}
\includegraphics[width=1.0\columnwidth]{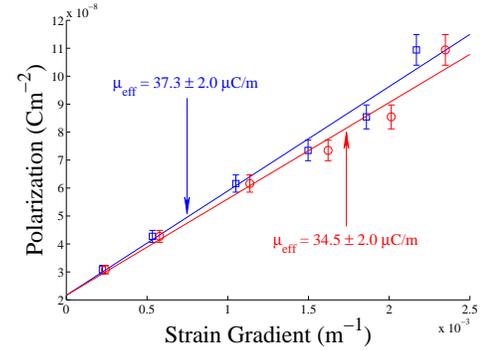}
\caption{The effective flexoelectric coefficients, $\mathrm{\mu_{eff}}$, of $\alpha$-phase PVDF films measured at room temperature; the line with the ``circle maker'', whose slope is $\mathrm{\mu_{eff}=34.5\pm2.0 \mu C/m}$, represents the measured results by using our method; the line with the ``square maker'', whose slope is $\mathrm{\mu_{eff}=37.3\pm2.0 \mu C/m}$, represents the measured results by using the Ma-Cross's method.}
\label{fig4}
\vspace{-0.25in}
\end{center}
\end{figure}

If we further assume that the deflection of the effective beam system under bending conditions is infinitesimal, then the relationship between its strain in the X axis direction and its deflection $w(l)$ in the Z axis direction can be written as \cite{bisplinghoff2002}
\begin{equation}
S_{11}=-z\frac{\partial^{2}w(l)}{\partial l^{2}},
\label{flexo5}
\end{equation}
where $S_{11}$ is the applied strain; ``1" represents the X axis, $l$ is the variable on the X axis. Similarly, we also use ``2" and ``3" to represent the Y axis and the Z axis, respectively. The minus sign given in Eq.\ (\ref{flexo5}) only indicates whether the top surface of the beam system is in the tensile state ($S_{11}>0$) or the compressive state ($S_{11}<0$). The absolute value of the gradient of the strain along the Z axis direction is then given by
\begin{equation}
\left|\frac{\partial S_{11}}{\partial z}\right|=\frac{\partial^{2}w(l)}{\partial l^{2}}.
\label{flexo6}
\end{equation}

According to the Kogan's model defined in Eq.\ (\ref{flexo1}), the linear relationship between the elastic strain gradient and the electric polarization of the PVDF film under the bending condition, shown in Fig. 2(c), can be written as
\begin{equation}
P_{3}=\mu_{eff}\frac{\partial S_{11}}{\partial z},
\label{flexo7}
\end{equation}
where $P_{3}$ represents the induced electric polarization along the Z axis direction; $\mu_{eff}$ is defined as the ``{\it effective flexoelectric coefficient}". $P_{3}$ can be further calculated by using the following equation:
\begin{equation}
P_{3}=\frac{I}{2\pi fA},
\label{flexo8}
\end{equation}
where $I$ is the induced alternating current; $f$ is the driving frequency of a dynamic loading tester (DLT) diagrammatically shown in Fig. 3(a), and $A$ represents the area of the top electrode, which is demonstrated in Fig. 3(b). In our experiment, the film samples we studied were neither stretched nor poled so that we could exclude possible extrinsic factors from our flexoelectric measurement; besides, such samples can be approximately assumed to possess a point group symmetry $\infty\infty m$ since the $\alpha$-phase domains in those samples are randomly oriented. Thus we can re-define $\mu_{eff}$ as: $\mu_{eff}=\nu\mu_{11}+(1+\nu)\mu_{12}$, here $\nu$ represents the Poisson's ratio, and both $\mu_{11}$ and $\mu_{12}$ are the flexoelectric coefficients in matrix notation \cite{newnham2005}. Since determining $\mu_{11}$ and $\mu_{12}$ needs tedious measurement of the applied strain gradient and the induced electric polarization and their explanations also need more space, which is not suitable for a brief research report, we will not discuss the details of the measurement here. In this study, we mainly investigate the effective flexoelectric responses of $\alpha$-phase PVDF films under bending conditions.

There is still a concern about the Ma-Cross method, i.e., whether the assumption that the effective beam system is undergoing a natural vibration is appropriate or not. This is because Ma and Cross actually used the forced vibration not the natural vibration to generate the bending movement of the cantilever beam with one end clamped in their studies \cite{ma2001a,ma2001b,ma2002,ma2003,ma2005}. Therefore, the deflection $w(l)$ defined by Eq.\ (\ref{flexo4}) should be re-examined, which is another research objective of this Letter. We consider a clamped effective beam system under an end load shown in Fig. 2(d); its static deflection, $y(l)$, is given by \cite{gere1997}
\begin{equation}
y(l)=A_{2}l^{2}\left(3L-l\right),
\label{flexo9}
\end{equation}
where $A_{2}=-F/6EI$; $F$ represents the end load, $E$ and $I$ are the modulus of elasticity and the moment of inertia of the clamped effective beam system, respectively; $l$ and $L$ are defined in Fig. 2(d). If we let $F$ vary slowly, $y(l)$ will then become quasi-static, which means that the bending movement of the beam system shown in Fig. 2(d) could be generated by a forced low-frequency vibration. Similarly, $A_{2}$ can also be determined from the boundary condition, i.e., the measured displacement in the Z axis direction of the effective beam system. Both the direction of $y(l)$ and the Z axis are shown in Fig. 2(d). When $y(l)$ is very small, the relationship between the gradient of the generated strain along the Z axis direction and the deflection $y(l)$ can be written as

\begin{equation}
\left|\frac{\partial S_{11}}{\partial z}\right|=\frac{\partial^{2}y(l)}{\partial l^{2}}.
\label{flexo10}
\end{equation}

Our flexoelectric measurement is summarized as follows: the aforementioned forced low-frequency vibration of the beam system is driven by the DLT shown in Fig. 3(a), which is modified from a commercial piezo actuator; a function generator (FG) is used to provide an AC sinusoidal signal with a frequency of $f = 0.5 \mathrm{Hz}$ to an amplifier to control the periodic movement of the DLT and to a lock-in amplifier (LIA) as the reference signal. The dynamic bending of the beam system is controlled via the DLT, and its displacement or deflection in the Z axis direction is detected by using a MicroStrain DVRT probing sensor; the measured displacement values are used to calculate the strain gradient along the Z axis direction via Eqs.\ (\ref{flexo4}) and (\ref{flexo6}) (by using the Ma-Cross method) or via Eqs.\ (\ref{flexo9}) and (\ref{flexo10}) (by using our method). Finally, using Eq.\ (\ref{flexo8}), we can obtain the induced polarization, $P_{3}$; here $I$ is collected by using the LIA; $l$, $L$, and $A$ are shown in Fig. 3(b). Under the same physical conditions and at room temperature, we measured the flexoelectric responses of $\alpha$-phase PVDF films under bending conditions. We obtained approximate values of the effective flexoelectric coefficients, $\mathrm{\mu_{eff}=34.5\pm2.0 \mu C/m}$ (via our method) and $\mathrm{\mu_{eff}=37.3\pm2.0 \mu C/m}$ (via the Ma-Cross's method). The linear relationships between the induced polarization values and the applied strain gradient values are shown in Fig. 4.

We have to emphasize that, in both the Ma-Cross's method and our method, it is critical to keep the generated deflection as small as possible since Eq.\ (\ref{flexo6}) and Eq.\ (\ref{flexo10}) can only be used to calculate the strain gradient when the deflection is infinitesimal. So, in theory, the testing point, the spot covered by the top electrode shown in Fig. 3(b), on the beam system should be chosen near its clamped end. However, a perfectly rigid clamped condition for a cantilever beam can never be achieved in practice; an imperfect clamped condition often makes the lock-in detection very difficult. Therefore, the testing point for such flexoelectricity measurement should be carefully chosen to reduce the inevitable errors caused by the imperfect clamped condition. In our experiment, we chose the spot on the beam system, where the ratio $l/L\approx1/5$ shown in Fig. 3(b), as our testing point.

In conclusion, a giant direct flexoelectric response in $\alpha$-phase PVDF films under bending conditions has been observed; we also found that the induced flexoelectric polarization is indeed linearly proportional to the applied strain gradient in our samples. It is perhaps interesting to compare the giant flexoelectric phenomena observed in polymers with that in solid dielectrics. Ma and Cross observed a giant flexoelectric phenomenon, $\mathrm{\mu_{12}\approx100 \mu C/m}$, in a barium strontium titanate ($\mathrm{Ba_{0.67}Sr_{0.33}TiO_{3}}$ or BST) composition \cite{ma2002}; in their studies, the BST composition is in the paraelectric state just above its Curie temperature, and, therefore, its relative permittivity value is very large, $\varepsilon_{r}\approx16000$ \cite{ma2002}. According to the Tagantsev's model, it is not surprised that giant flexoelectricity can be observed in such a solid material. However, the average relative permittivity value of our $\alpha$-phase PVDF samples is only $\varepsilon_{r}\approx11$, which is three orders of magnitude smaller than that of the BST. This indicates that the Tagantsev's model might not be applied to the flexoelectric behavior of polymers. Finally, we also verified that the measured flexoelectricity results by using the Ma-Cross method and our method are in good agreement when the frequency of the periodic external driving force is low.

\begin{acknowledgments}
We would like to thank Solvay Solexis for providing PVDF powders to us. The research reported in this Letter was sponsored by the State University of New York at Buffalo.
\end{acknowledgments}
%
%
%
%

%
%
%
\end{document}